# Automated Discovery of Elementary Chemical Reaction Steps Using Freezing String and Berny Optimization Methods

Yury V. Suleimanov[1,2,*] and William H. Green[2]

[1]*Computation-based Science and Technology Research Center, Cyprus Institute, 20 Kavafi Street, Nicosia 2121, Cyprus*

[2]*Department of Chemical Engineering, Massachusetts Institute of Technology, Cambridge, Massachusetts 02139, Unites States*

*\* Email: ysuleyma@mit.edu*

We present a simple protocol which allows fully automated discovery of elementary chemical reaction steps using in cooperation single- and double-ended transition-state optimization algorithms – the freezing string and Berny optimization methods, respectively. To demonstrate the utility of the proposed approach, the reactivity of several systems of combustion and atmospheric chemistry importance is investigated. The proposed algorithm allowed us to detect without any human intervention not only 'known' reaction pathways, manually detected in the previous studies, but also new, previously 'unknown', reaction pathways which involve significant atom rearrangements. We believe that applying such a systematic approach to elementary reaction path finding will greatly accelerate the possibility of discovery of new chemistry and will lead to more accurate computer simulations of various chemical processes.

# Introduction

Currently, new chemical understanding is driven mainly by the experimental discovery of new compounds and their reactivity. This experimental work is expensive and time-consuming, but the payoff for discovering new chemical reactions is enormous in terms of both fundamental understanding and practical engineering. Examples of important new reactions discovered in recent decades include many advances that followed from the discoveries of olefin metathesis[1] and click chemistry.[2] Theoretical, algorithmic and computational advances in quantum chemistry over the last two decades have brought it to the stage where it is widely applied for qualitative interpretation of such new experimental findings, and increasingly used to construct detailed quantitative descriptions of chemical reaction networks.

The main parameters in chemical kinetic models – equilibrium constants and rate coefficients – are increasingly derived from quantum mechanical calculations on stationary points (local minima and first-order saddle points) along the potential energy surface (PES). Local minima, i.e. reactants, intermediates, and products, are generally easy to characterize due to simple bonding and because the negative of the gradient along the PES always points downhill. Several robust algorithms were developed for identifying local minima.[3] Saddle points, which correspond to transition states, remain a challenge for systematic characterization due to non-standard variations in reactive centers and because a transition structure optimization must step uphill in one direction and downhill in all other orthogonal directions. At present, most saddle points are found based on human intuition, derived primarily from experience on analogous reactions: the human provides a very good initial guess at the saddle point geometry, and then various algorithms (so-called single-ended) are used to refine the initial guess geometry to find the true saddle point.[4-8] This has typically been applied to cases where humans expect a reaction to connect the reactant(s) and product(s), and so it also rarely discovers anything surprising. This conventional approach turns out to be quite good for finding "expected reaction" saddle points, but rarely discover anything that is unexpected. As an alternative to human input, the initial guess structure for the saddle point can be generated by methods (so-called double-ended searches) which reconstruct the reaction path between known reactant and product sides.[9-13]

Most large chemical kinetic models are constructed in ways which do not allow any unexpected reactions to be included; this is true whether the model is constructed by hand or using automatic mechanism generators.[14] Incorporation of even a simple new reaction in chemical kinetic models can significantly alter these models since almost certainly it will be the first member of a new large family of related reactions. For instance, several unexpected low-barrier reactions for JP-10 pyrolysis[15] and for decomposition of the ketohydroperoxides,[16] which are critical intermediates in alkane autooxidation and ignition chemistry, were recently discovered in our group. Including the new unexpected reactions and analogous new reactions in the kinetic models significantly improved the reliability of the kinetic model predictions.[17]

All this experience suggests that if we search for them systematically, we can expect to find many "unexpected" reactions which are omitted from existing chemistry models, and that incorporating these new reactions will significantly improve the reliability of the model predictions. In the present work, we propose a systematic approach for discovery new chemical reactions with minimum human effort. The method advances recent work on saddle point optimization algorithms, by making use of both single- and double-ended methods in a fully automated procedure. Among a great variety of methods,[4-13] we employ a combination of the Berny optimization algorithm as implemented and improved by Schlegel[6-8] and the freezing string method[13] as both methods were readily available to us at the time we started this project. We find that this combination works fairly well for several systems of combustion and atmospheric chemistry importance. We were able to detect not all 'known' reaction pathways, manually detected in the previous studies, but also new, previously 'unknown', reaction pathways, which involve significant atom rearrangements. The present work is complementary to recent developments in computational chemistry[14,18] which demonstrate a tendency of switching from a traditional secondary role of theory in interpreting experimental findings in chemistry to a powerful predictive tool for analysis of chemical reactivity, discovery of chemical reaction pathways, and generation of reaction mechanisms prior to any experiments.

**Theory**

**Matrix Representation of Species and Reactions**

There are different ways to describe mathematically compounds presented in a system.[19-26] In the present algorithm, we chose the Bond Electron (BE) matrices[19,20] as one of the most vivid and convenient representations. A BE matrix **R** for a compound with $n$ atoms $r_1, r_2, ..r_n$ is a square matrix of dimension $n$. The $i^{th}$ row of this matrix corresponds to electron configuration of atom $r_i$. The $R_{ij}$ element represents the number of covalent bonds between atoms $r_i$ and $r_j$ and matrix **R** is symmetric, i.e. $R_{ij} = R_{ji}$. The diagonal elements $R_{ii}$ represent the number of free valence electrons which are not involved in bonds. The sum over the entries of the $i^{th}$ row or column of a BE-matrix is the number of valence electrons which belong to the atom $r_i$.

The total matrix of the initial reactive system $\mathbf{R}_{sum}$ is depicted by a block matrix, each block representing one compound. A chemical reaction is the conversion of the system by the redistribution of valence electrons. It is represented by transformation of the initial matrix $\mathbf{R}_{sum}$ by the addition of the reaction matrix $\mathbf{A}$ to yield a product total matrix $\mathbf{P}_{sum}$ with altered connectivity. The reaction matrix $\mathbf{A}$ is also symmetric and its elements represent the appearance (positive value) or the disappearance (negative value) of the localized valence electrons. This matrix should conserve the total number of electrons in the system, i.e. $\sum_{i,j} a_{ij} = 0$. Usually in elementary reactions $|a_{ij}| \leq 2$, since the change by two implies a significant chemical process, for instance, formation or rupture of a double bond in a single elementary step.

After applying the reaction matrix $\mathbf{A}$, the product matrix $\mathbf{P}$ can then be rearranged to form a block structure to represent the product molecules. If the product matrix satisfies the mathematical and chemical constraints, the reaction pathway is considered to be feasible and is added to the list of channels for further thermochemistry calculations.

**Reducing the Size of the Search Space**

*Thermochemistry Calculations.* Even for a small hydrocarbon system, the number of reaction pathways which can be generated using the BE matrix representation described above can be massive, increasing exponentially with the number of carbon atoms. The thermodynamic reaction control principle can be implemented to narrow the search space: highly endothermic reactions are likely to be too slow to be important. The relative stability of the products can be estimated using their thermochemical properties, and reaction channels with less stable products can be eliminated. However, on-the-fly quantum-chemical calculation of thermochemical properties for hundreds (or even thousands) species using, e.g. density functional theory (DFT) methods, can also be computationally very demanding. One efficient alternative way to do so is to use group contribution methods that estimate the thermochemistry of a molecule based on the sub-molecular fragments present in the molecule. The Benson group additivity framework is such an example of a group contribution method that has proven to provide accurate estimates of the ideal gas thermochemistry for a large range of molecules. Benson's group additivity approach[27] divides a molecule into functional groups, and the contribution of each functional group to the overall thermochemistry is included. The group additivity enthalpy estimate is typically within a few kcal/mol[28] of the truth, sufficient to reliably separate very endothermic reactions from the rest.

*Generating 3D Geometries.* While DFT methods can, in principle be implemented to generate and optimize the starting and final 3D geometries, the large number of reaction channels can prohibit such calculations. A much faster route is to use a molecular mechanics force field. Though, in principle, any type of reasonable force field can be used, we choose Merck Molecular force-fields (MMFF94) developed by Merck Research Laboratories[29-33] which were specifically optimized for a wide range of organic chemistry calculations (from very small molecules to proteins). The method includes parameters from high-quality quantum calculations (rather than experimental data) for a wide range of atom types including those presented in the examples studied in this work.

**Locating Transition States**

As mentioned in Introduction, methods for transition state structure search can be classified as single-ended and double-ended.[4,5] The single-ended methods refine an initial guess of the transition state to the exact answer by exploring the PES using local gradient and usually the second derivative information. The Berny saddle point optimization algorithm is one of these methods, which is widely used in *ab initio* calculations primarily due its reliability.[6-8] Berny geometry optimization starts with an initial guess for the second derivative matrix (Hessian) which is determined using connectivity derived from atomic radii and a simple valence force field.[7,8] The approximate Hessian matrix is improved at each point using the first derivatives and energies computed along the optimization pathway. The main input for such algorithm is a guess structure that should be located very close to the transition state in order to succeed. At present, initial guess geometries are usually found based on human intuition, and automated generation of good initial guess transition state structures is the main point of the procedure proposed here.

The double-ended methods start from the reactants and products and work from both sides to find the transition structure and the reaction path. These include the nudged elastic band method,[9,10] string method,[11,12] and freezing string method (FSM).[13] All these algorithms generate a sequence of configurations located between the reactant and products geometries. In the present algorithm we choose FSM as it does not require Hessian calculations and provides one of the most inexpensive algorithms for determining a path with a specified number of intermediate structures (nodes) connecting the known reactant(s) and product(s). FSM starts from both ends of the path, and adds nodes irreversibly until the two ends join. FSM partially optimizes intermediate structures orthogonally to the reaction path, and then freezes them before a new pair of points along the reaction path is added. The number of gradients required to locate a guess saddle point using FSM is rather small.

Since the reactant and product geometries, corresponding to initial and final minima, are known from the matrix representation, energy cutoff and force field optimization described above, automatic path-finding tools, such as the freezing string method, can characterize a reaction coordinate joining the end-points. The highest point on the pathway becomes a good initial guess for subsequent refinement of the transition structure using the single-ended method such as Berny algorithm.

**Summary of methodology and computational details**

All the methods just described can be combined in a systematic and automated procedure for identifying reaction pathways. It is summarized in the flowchart presented on Figure 1. Only initial reactant(s) geometry is required at the start. It is converted to the reactant matrix **R**. Different reaction matrices **A** are then generated, which break up to $N$ bonds and make up to $M$ bonds, where $N$ and $M$ are defined by the user. The number of possibilities rises combinatorially with $N$ and $M$. In the benchmark examples below, we constrain $1 < N < 4$ and $1 < M < 4$. We constrain $N$, $M < 4$ to reduce computational demands and $N$, $M > 1$ since most of those simple reactions are well understood. To reduce the computational effort we restricted the products to those which can be drawn with a Lewis structure with a fixed number of valence electrons for each element – 4 for carbon, 2 plus two lone pairs for oxygen, 1 for hydrogen, 3 plus a lone pair for nitrogen.

In the present work, we restrict all the elements $R_{ij}$ and $P_{ij}$ in the reactant and product matrices to be integers. This can potentially lead to issues for products with resonant structures. We also demand number of unpaired electrons on an atom not exceed 1. This is certainly ruling out some known species, e.g. biradicals such as methylene and O($^1$P), but we can always relax this restriction and include them in future studies. Since BE matrices are invariant with certain permutations, several BE matrices can correspond to chemically equivalent compounds. However, in order not to omit important reaction pathways, all combinations are taken into account initially. We remove any duplicates before the molecular mechanics and quantum chemistry calculations are run.

The product matrices **P** are obtained and converted to block structure by simple linear algebra operations. Thermochemistry calculations are used to narrow the search space - during this step 'known' and unstable products are detected. In the present study, a reaction is considered to be too endothermic to be interesting if the standard enthalpy of reaction (denoted as $\Delta H_r^0$ in Tables 2-5) is higher than 20 kcal/mol. These calculations are performed using the Reaction Mechanism Generator (RMG) software package developed in our group.[14]

The next step is automated saddle point search for 'unknown' reaction pathways which lead to stable products. For quantum chemistry calculations, geometries of reactant(s) and product(s) are required. They are generated by converting corresponding blocks of the **P** matrices to SMILES strings (standard format for molecule mechanics calculations) and quick force field optimization using academic version of the MarvinBeans package from ChemAxon.[34] Special care is taken of the conversion between different molecule representation formats, eliminating duplicate channels. In the present study, we employ the MMFF94 force field to determine the geometry of the products.

The BE matrix representation is identical for all conformers. However, the distinct 3d structure of conformers strongly affects the properties and the reactions of molecules. In order to take into account conformational effects, we used the MarvinBean's Conformer Plugin to generate a series of stable 3d structures (up to three in the present study), i.e. conformers for each compound. If more than one product is formed in the reaction channel, their structures were first aligned to maximum coincidence with the reactant structure in non-mass weighted Cartesian coordinates and then the length of the vector between the centers of mass of the products was increased in order to avoid overlaps between atoms in different product molecules.

For the determination of all the transition states we used the FSM implemented in the Q-Chem software package with default spacing parameters (20 nodes, 6 perpendicular gradients per node).[35] In order not to repeat computations on duplicate structures, all input structures are compared prior to initiating FSM calculations and identical input files were eliminated. However, chemically equivalent channels with distinct 3d geometries were not excluded from the procedure in order not to omit conformers or atom arrangements which could potentially lead to low-energy reaction pathways. For channels where FSM found an apparent single barrier along the reaction path (without any restrictions to the energy barrier height), these calculations were followed by transition state search using the Berny optimization algorithm implemented in the Gaussian 03 software package[36] starting from the FSM geometry. For each detected saddle point, we also perform intrinsic reaction path (IRC) calculations which go downhill from the transition state to see which two minima it is connected to and to verify whether this saddle point corresponds to the reaction from *R* to *P*.

Geometry optimization and reaction path analysis for the examples herein were performed using M062X/6-311++G* level of theory, while single point energies at the stationary points were obtained using CCSD(T)/6-311++G* level of theory. Thermochemistry of the detected reactions was computed using CBS-QB3 level of theory. Note that the proposed procedure is not dependent on the use of any particular software package or level of theory. Our choice of such parameters as quantum chemistry methods, basis set, force fields, FSM parameters, number of conformers, method for orientating product structures should be considered as reasonable first attempts which can be improved in the future as the community gains more experience.

Each set of quantum chemistry calculations – structure optimization, energy calculation and the FSM calculation – is performed independently in parallel to exploit the full potential of supercomputers. All the computations are automated, to reduce time-to-completion and also to reduce bias related to manual intervention to help the computer find a saddle point, and which so tend to find only "human-expected" known reactions. In order to demonstrate how such fully-automated procedure can accelerate the discovery of new chemistry, we use it to study several systems of combustion, oxidation and atmospheric chemistry importance which span a reasonable wide space of chemical reactivity.

**Results**

For our first case study we have chosen chemistry of $\gamma$-ketohydroperoxide ($HOOCH_2CH_2CHO$). Ketohydroperoxides are the main source of radicals during the first stage of low-temperature ignition, and are also formed in liquid phase oxidation. Combustion parameters, such as ignition delay, are very sensitive to the details of their chemistry.[16] Most previous combustion models assumed ketohydroperoxides do only one reaction: forming radicals by breaking the weak O-O bond. However, new chemical reactions of ketohydroperoxides which transform them into cyclic peroxides and then acids and carbonyl molecules has been recently characterized[16] in our group (so-called "Korcek reaction"),[37] suggesting that chemistry of ketohydroperoxides is more complicated. To verify the generality of the proposed algorithm, we also studied other structures - 1,2-dioxolan-3-ol (intermediate cyclic peroxide in the Korcek mechanism),[16] ethylene ozonide (an intermediate in reactions between the Criegee intermediate and carbonyl compounds important in atmospheric ozonolysis models)[38] and their nitrogen counterparts N-(hydroperoxymethyl) formamide, 1,2,4-dioxazolidin-3-ol, and 1,2,4-dioxazolidine – model systems for oxidation chemistry of nitrogen-containing oligomers, important components of lubricants and fuel. We also analyzed chemistry of penta-1,4-diene – a model system for an important functional group in polyunsaturated lipids. For each initial structure, Table 1 shows the number of product channels generated using the BE representation. The results of thermochemistry analysis using Benson group additivity approach of all product channels are also given in Table 1. All reaction pathways identified are summarized in Figures 2-5 and Tables 2-5. The Cartesian coordinates of all the automatically detected saddle points are summarized in the Supporting Information.

For the first case study, $\gamma$-ketohydroperoxide, Table 1 shows that almost 500 product channels, which break and make 2 or 3 bonds, are possible from this small molecule. (Note that this includes distinct reactions which form the same products. Without repeated channels, this number reduces to 109.). Half of the channels are exothermic and almost 70 % of the channels have $\Delta H_{rxn}$ < 20 kcal/mol. At the same time, only one (!) product channel (1,2-dioxolan-3-ol) is reached by a "well-known" reaction type available in our current RMG database – the first step of the Korcek mechanism forming an intermediate cyclic peroxide.[16] As shown in Figure 2 (Reaction (1)), our simple algorithm was able to locate efficiently transition states for this 'known' reaction though it involves a significant rearrangement of atoms including cycle formation and hydrogen transfer between two oxygen atoms. Figure 2 also shows that the automatic search found transition states for 5 additional reaction pathways. Note that one of these channels leads directly to the products of the Korcek mechanism - Reaction (4) on Figure 2. However, Table 2 shows that the energy barrier (denoted as $\Delta E^{\#}$) for this channel is higher by almost 20 kcal and therefore, rather than going directly over this high mountain pass, sequential reactions via an intermediate valley (the cyclic peroxide) are expected to be the fastest reaction path. This channel was located while searching for a reaction path to a different product (see Comments in Table 2). This example illustrates that our saddle-point search procedure cannot guarantee convergence to the desired transition state. As a result of its non-iterative nature, the FSM cannot guarantee conservation of the exact IRC profile and therefore further optimization of the guess structure (the highest energy node) with Berny optimization can lead to structures which do not correspond to the initial reaction path. We also occasionally converged to saddle points which correspond to different initial reactants. More importantly, the majority of our reaction path searches do not converge to a saddle point. This typically occurred because the energy profiles from the FSM calculations for these paths exhibited several very high barriers – not the single elementary-step transition state we were looking for – and therefore the Berny-type TS search was not performed. In these cases, it is possible that no single saddle point exists that connects reactant and product(s), or just that with our choices of parameters and options the FSM is not robust enough to find one. Clearly improved methods are needed.

Upon examination of the reactivity of 1,2-dioxolan-3-ol – our second case study - we confirmed that fragmentation of the cyclic peroxide leads to two possible pairs of acid and aldehyde products (Reactions (24) and (25) in Figure 4) previously characterized in Ref. 16. We were also able to locate the reverse pathway which leads to the initial molecule (Reaction (22) in Figure 4) confirming that the proposed procedure is rather sustainable. The fact that the fully automated search reproduces the results of the previous manual studies of the Korcek mechanism – and also discovers a half-dozen additional transition states to products not found by that manual search – is encouraging.

Another interesting example is the chemistry of the Criegee intermediate with formaldehyde (third case study). These two species quickly recombine forming the more stable secondary ozonide intermediate (ethylene ozonide)[38] whose reactivity was investigated in the present work. All three reaction pathways identified for ethylene ozonide are presented in Figure 3 (Reactions (15)-(17)) and Table 3. In accordance with the previous manual search [38], we find that two low energy pathways for ethylene ozonide decomposition are formation of hydroxylmethyl formate by breaking the weak O-O bond (the lowest energy saddle point) and formation of formaldehyde and formic acid. In addition to the reactions found in Ref. 38, we also find that ethylene ozonide can decompose to formaldehyde and oxirane with similar barrier height.

We also repeated the calculations for similar structures with nitrogen-substituted analogues: N-(hydroperoxymethyl) formamide, 1,2,4-dioxazolidin-3-ol, and 1,2,4-dioxazolidine. Table 1 shows that the chemistry of these N-containing compounds is even less known – no unimolecular channels with 2 and 3 bonds breaking/forming were found in the RMG database. However, we were able to identify and characterize several reaction pathways using our automated procedure. The results are summarized in Tables 2-4 and Figures 2-4. They show that the chemistry for these molecules is similar to the first three structures. For example, Reaction (7) is similar to the first step of the Korcek mechanism (Reaction (1)). However, replacing one $CH_2$ group with an NH group significantly enriches the chemistry. For example, Reaction (9) involves two H-transfers (transition state shown in Figure 2) and Reactions (11) and (12) form 3 products from one reactant.

Table 2 shows that in the case of N-(hydroperoxymethyl) formamide, half of the identified saddle points do not correspond to the intended product(s) generated by the BE matrix representation. Due to initial open-chain structure, all these cyclization reactions require significant rearrangement of several atoms, and apparently FSM has trouble with these cases. Similarly, the algorithm failed to find several cyclization transition states of the $\gamma$-ketohydroperoxide. At the same time, all saddle points detected for more constrained cyclic reactants 1,2,4-dioxazolidine and 1,2,4-dioxazolidin-3-ol are for the intended products (see Tables 3 and 4).

As a final example, the chemical reactivity of 1,4-pentadiene is investigated (see Table 5 and Figure 5). Due to many possible single and double hydrogen atom transfers, many product channels are possible in this system. Table 1 shows that total number of channels exceeds seven hundred. However, out of this number, only 8 saddle points were located and 3 of these correspond to known reaction types in the RMG database. All structures were connected to the initial molecule by a large barrier (higher than 60 kcal/mol). Nevertheless, the ability of our algorithm not only to reproduce the existing 'known' reaction pathways but also to identify 5 new reaction pathways for this simple molecule is very promising.

## Conclusions

The algorithm proposed here for finding elementary reaction paths possesses several important features: it requires no human intervention and it does not require any information on the chemical reactivity of the reactant(s). In order to detect saddle points it generates a series of product(s) and tries to find pathways that connect the product(s) to the initial structure - this would be tedious to do without automation. It is much more computationally efficient (i.e. requires fewer energy/gradient calculations) than methods based on *ab initio* molecular dynamics, since it rapidly finds and focuses on the structures near the saddle point connecting the reactant with designated products. The present examples show that our automated algorithm is able to detect not only previously detected reaction pathways but also several new, previously unknown, types of reactions. Our constraint that allows only two or three bonds to be formed and broken in the same elementary step worked well for the present examples, keeping the number of possible product channels manageable (< 1000 possible reaction paths per reactant) but still allowing us to discover a lot of new chemistry. The present results also show that detecting thermodynamically feasible products does not guarantee kinetic feasibility. The degree of success in locating unexpected kinetically favorable elementary steps is rather low – a significant number of thermodynamically stable structures either are connected to the initial structure only through multiple elementary steps or via chemically infeasible routes. In future, some of these routes which are the most obviously unlikely, such as breaking the C=O bond, might be eliminated in order to focus resources on discovering transition states for more likely reactions. Another possibility is to freeze certain bonds and angles that are expected to be unreactive in order to reduce the number of conformers and matrices dimensionality. However, such limitations should be imposed with caution as they may prohibit detecting unexpected important reaction pathways!

The success of the exact search for saddle point relies on the ability to generate very accurate corresponding guess structures from a double-ended method, such as the freezing string method (FSM) employed in the present study. FSM can be considered as one of the fastest double-ended methods. However, FSM is not guaranteed to converge to the reaction path of interest and therefore convergence to the correct saddle point cannot be rigorously guaranteed.[13] For each detected transition state we performed the intrinsic reaction coordinate (IRC) integration calculation to recover both reactant and product sides. In several cases, we found that the Berny optimization of the highest node along the reaction path generated by FSM leads to wrong transition states, which do not connect the specified reactant and product states. Therefore, a good initial guess of the transition state remains one of the major challenges in the use of such algorithms for complex systems.[39]

As an alternative to FSM, its predecessor, the growing string method (GSM), an iterative algorithm that connect the reactant and the product via the IRC, can be implemented. Recent advances in the growing string method have allowed it to locate the exact transition state for a given reaction path.[39,40] A procedure for automatic reaction path finding which is based on GSM has been recently proposed.[41] However, while formally very attractive, the GSM is substantially more computationally expensive compared to the cost of FSM calculation (by up to a factor of 10), and in this work the FSM calculations are the limiting computational step in the algorithm. Moreover, even GSM can fail to locate a saddle point for a single elementary step as no double-ended string method is perfectly reliable. An important point of the present testing is that FSM is able to discover several important transition states recently discovered by manual searches.[16,38] However, future improvement of reliability as well as reducing computational costs of double-ended methods is obviously required.

In future, we plan to couple the automated approach for finding the 'unexpected' reactions with existing automated mechanism-generation software packages for 'expected' reactions, such as the Reaction Mechanism Generator (RMG).[14] The combination would efficiently generate much more comprehensive and reliable chemical kinetic models than is possible with existing model-construction techniques.

## Acknowledgements

Y.V.S. and W.H.G. gratefully acknowledge financial support from the Air Force Office of Scientific Research, through grant FA9550-13-1-0065. The views expressed here are solely those of the authors, and are not endorsed by the sponsor. This research used resources of the National Energy

Research Scientific Computing Center, a DOE Office of Science User Facility supported by the Office of Science of the U.S. Department of Energy under Contract No. DE-AC02-05CH11231.## Associated Content

### Supporting Information

The saddle point structures for the reactions discussed in this paper are freely available at http://pubs.acs.org

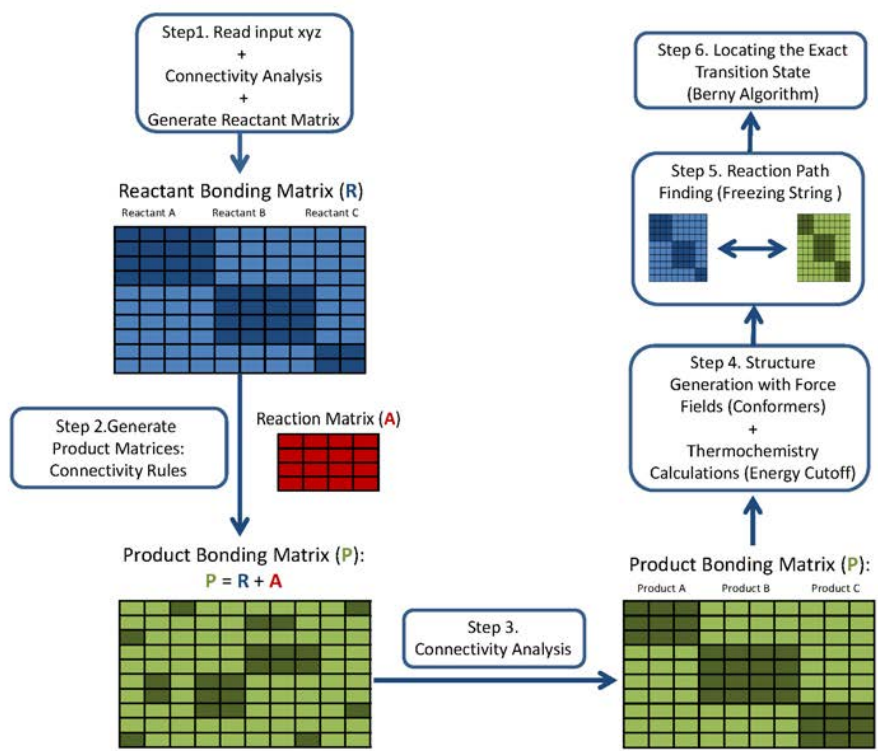

**Figure 1.** Flowchart for automated discovery of elementary chemical reaction steps using molecular mechanics and quantum chemistry calculations. See "Theory" section for details of each step.

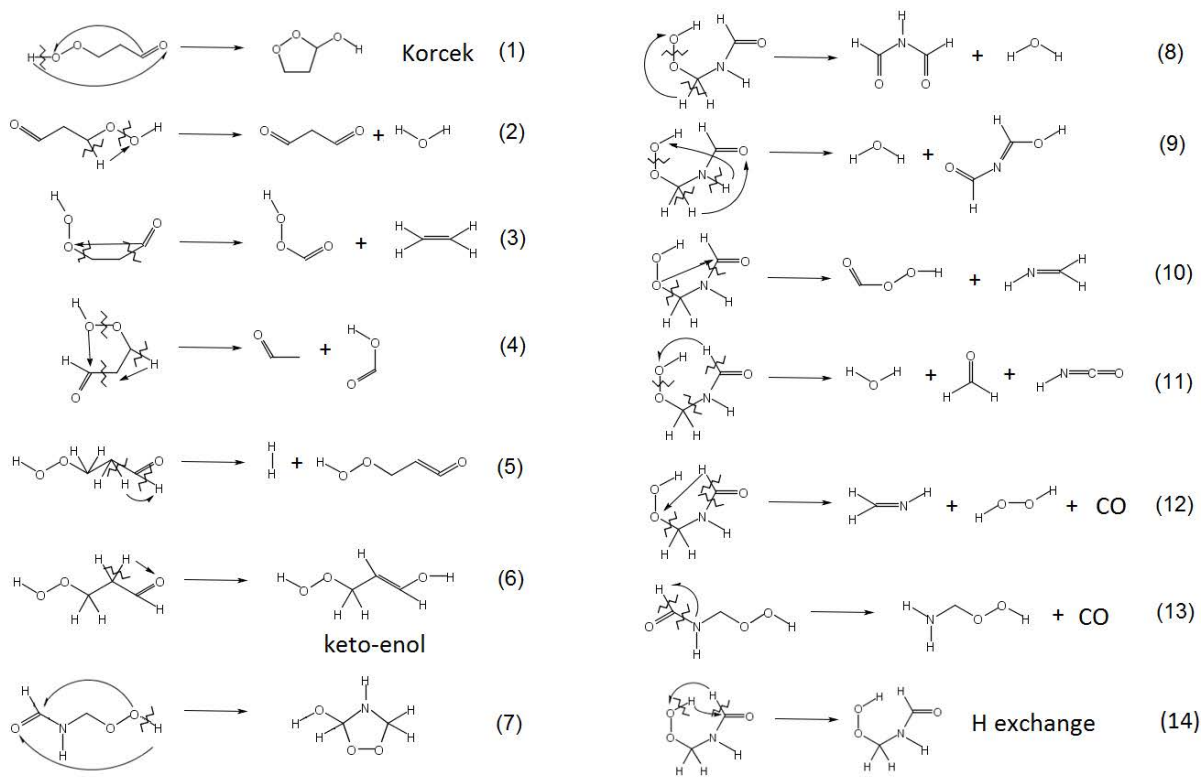

**Figure 2.** Automatically identified unimolecular reactions of γ-ketohydroperoxide (OOCCC=O) (1-6) and N-(hydroperoxymethyl) formamide (OOCNC=O) (7-14). Reactions (3), (4), (9)-(12) are unusual and unexpected.

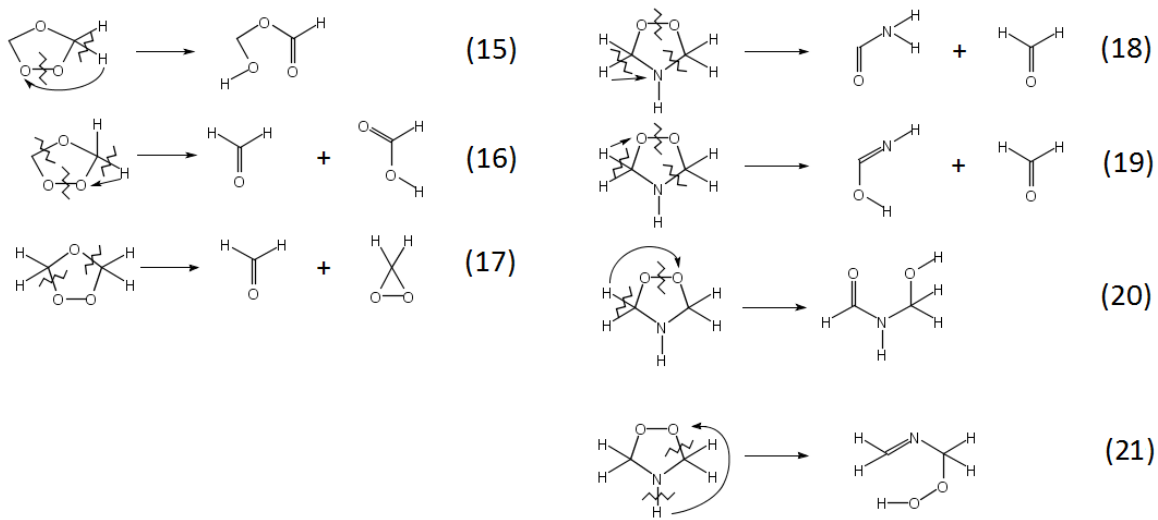

**Figure 3.** Automatically identified unimolecular reactions of ethylene secondary ozonide (O1COOC1) (15-17) and 1,2,4-dioxazolidine (N1COOC1) (18-21). Reactions (17) and (21) are particularly unusual and unexpected.

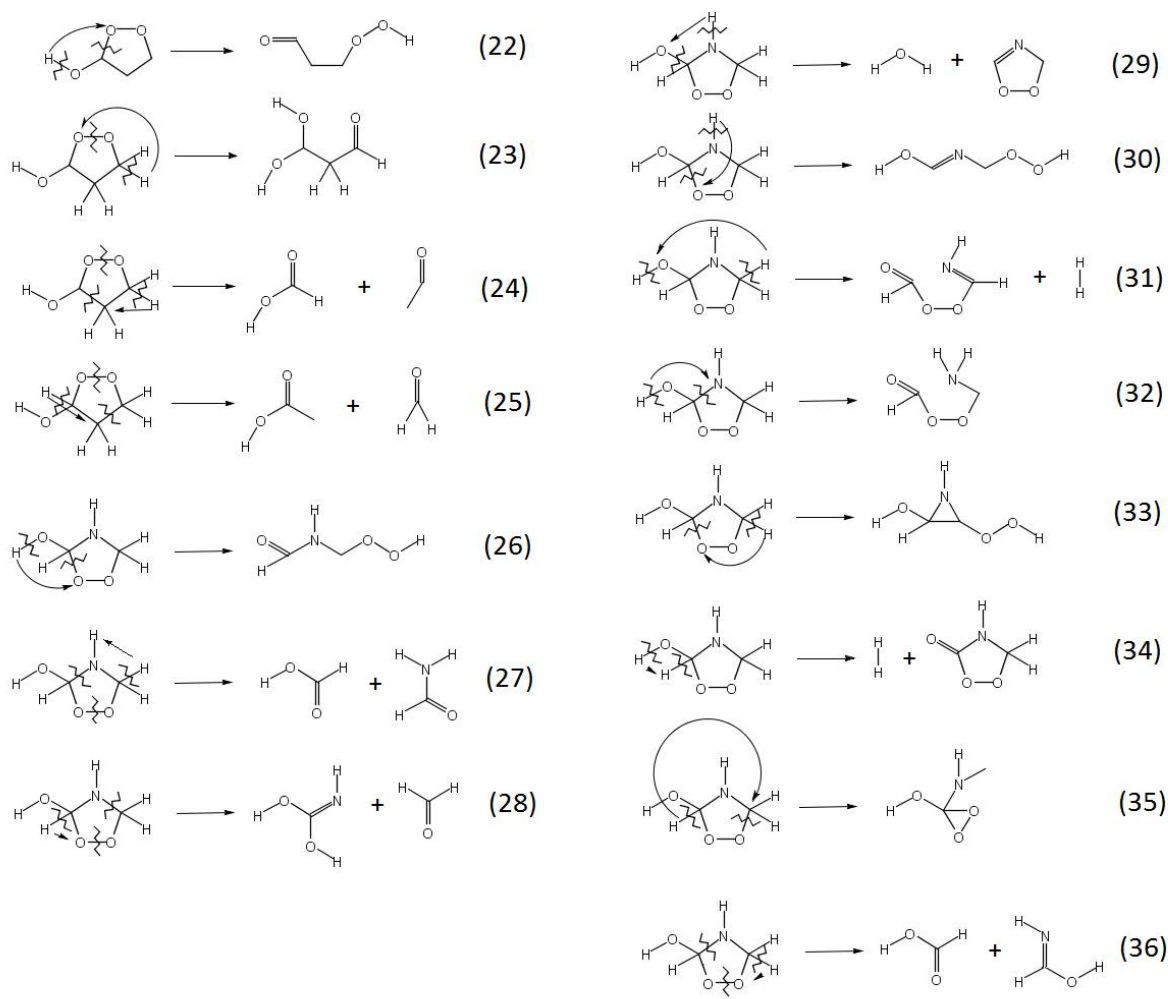

**Figure 4.** Automatically identified unimolecular reactions of 1,2-dioxolan-3-ol (OC1OOCC1) (22-25) and 1,2,4-dioxazolidin-3-ol (OC1OOCN1) (26-36). Reactions (30)-(33) and (35) are particularly unusual and unexpected.

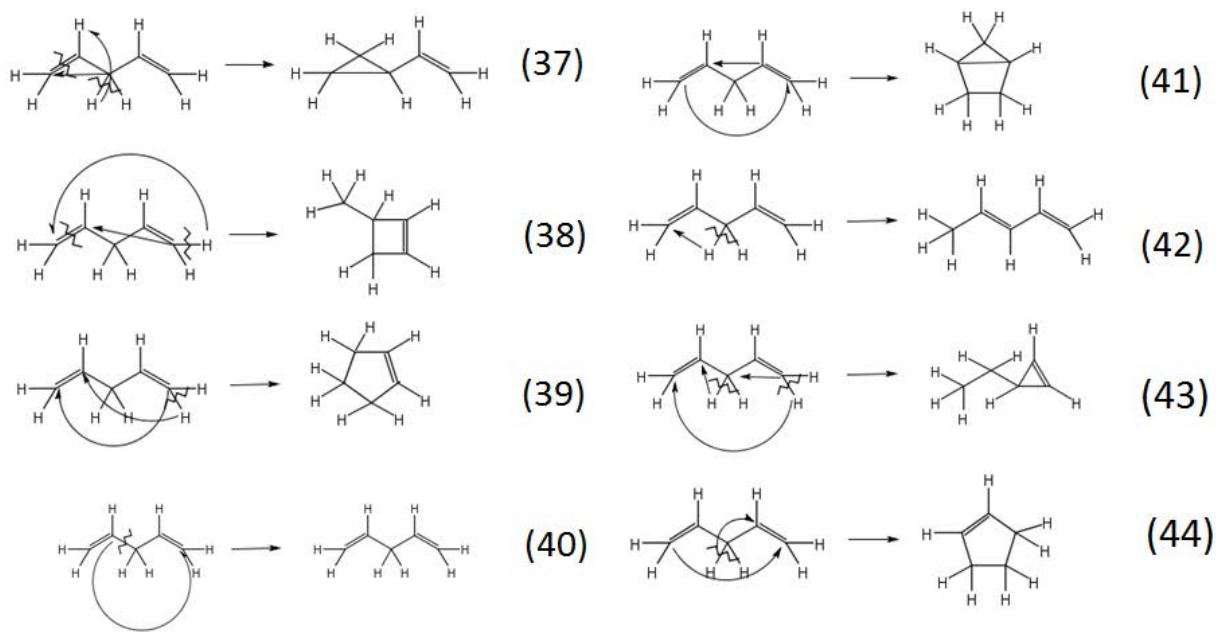

**Figure 5.** Automatically identified unimolecular reactions of 1,4-pentadiene (C=CCC=C) (37-44). Reactions (38)-(40) and (43) are particularly unusual and unexpected.

**Table 1.** Number of elementary steps investigated for different initial reactants.

| Initial reactant (SMILES) | Total number of product channels identified | Number of low energy channels | | Number of automatically identified transition states (versus existing RMG database)[c] |
|---|---|---|---|---|
| | | $(\Delta H_r^0(298K) \leq$ 0kcal/mol) | $(\Delta H_r^0(298K)$ $\leq 20$kcal/mol) | |
| γ-ketohydroperoxide (OOCCC=O) | 77[a] 415[b] | 39[a] 204[b] | 53[a] 276[b] | 6 (RMG database:1) |
| N-(hydroperoxymethyl) formamide (OOCNC=O) | 62[a] 293[b] | 22[a] 84[b] | 32[a] 123[b] | 8 (RMG database:0) |
| ethylene ozonide (O1COOC1) | 42[a] 82[b] | 23[a] 59[b] | 35[a] 80[b] | 3 (RMG database:0) |
| 1,2,4-dioxazolidine (N1COOC1) | 56[a] 190[b] | 35[a] 137[b] | 49[a] 159[b] | 4 (RMG database:0) |
| 1,2-dioxolan-3-ol (OC1OOCC1) | 90[a] 533[b] | 40[a] 233[b] | 64[a] 344[b] | 4 (RMG database:2) |
| 1,2,4-dioxazolidin-3-ol (OC1OOCN1) | 74[a] 378[b] | 43[a] 191[b] | 59[a] 275[b] | 11 (RMG database:0) |
| penta-1,4-diene (C=CCC=C) | 96[a] 652[b] | 45[a] 197[b] | 64[a] 341[b] | 8 (RMG database: 3) |

[a] Number of breaking bonds = 2; Number of forming bonds = 2

[b] Number of breaking bonds = 3; Number of forming bonds = 3

[c] from http://rmg.mit.edu

**Table 2.** Automatically identified unimolecular reactions of γ-ketohydroperoxide and N-(hydroperoxymethyl) formamide. In several cases, the computer was searching for a saddle point leading to intended products, but converged to a transition state for a different product channel. Mechanistic details are given in Figure 2.

| | Product(s) | $\Delta E^{\#a}$ | $\Delta H_r^0 (298\ K)^b$ | Comments |
|---|---|---|---|---|
| | 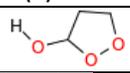 | | | |
| 1 | 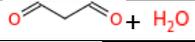 | 34.9 | -10.4 | Korcek reaction, see Ref. 16 |
| 2 | 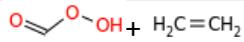 | 60.2 | -25.6 | |
| 3 | 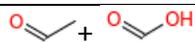 | 61.8 | 4.8 | |
| 4 | 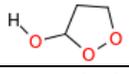 | 52.9 | -65.1 | Intended product: 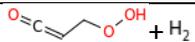 |
| 5 | 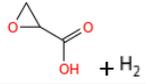 | 84.1 | 26.1 | Intended product: 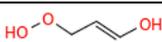 |
| 6 | 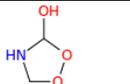 | 67.9 | 7.3 | Keto-enol |
| | 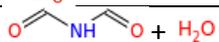 | | | |
| 7 | 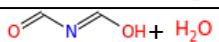 | 54.6 | 9.9 | Korcek-type reaction |
| 8 | 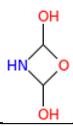 | 54.0 | -37.2 | |
| 9 | 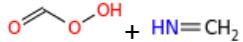 | 45.4 | -16.9 | Intended product: 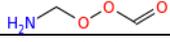 |
| 10 | 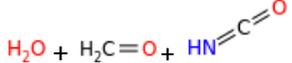 | 57.8 | 18.1 | Intended product: 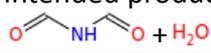 |
| 11 | 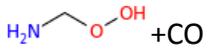 | 54.8 | -16.1 | Intended product: 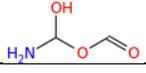 |
| 12 | $H_2C=NH + H_2O_2 + CO$ | 69.8 | 62.9 | |
| 13 | 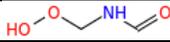 +CO | 83.3 | 12.6 | Intended product: $H_2N\diagdown O\diagdown OH\diagdown O$ |
| 14 | $HO\diagdown O\diagdown NH\diagdown O$ | 91.6 | 0.0 | H exchange |

[a] in kcal/mol; geometry optimization from Gaussian 03 using M062X level of theory, energy from Gaussian 03 using CCSD(T) level of theory. Basis set is 6-311++G* in both calculations.

[b] in kcal/mol; thermochemistry from Gaussian 03 using CBS-QB3 level of theory.

**Table 3.** Automatically identified unimolecular reactions if ethylene secondary ozonide (O1COOC1) and 1,2,4-dioxazolidine (N1COOC1). For mechanistic details see Figure 3.

| | Product(s) | ΔE[#a] | ΔH$_r^0$(298 K)[b] | Comments |
|---|---|---|---|---|
| | 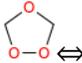 | | | |
| 15 | 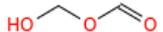 | 42.4 | -73.1 | See Ref. 38 |
| 16 | H$_2$C=O + 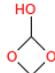 | 46.3 | -58.4 | See Ref. 38 |
| 17 | H$_2$C=O + 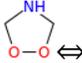 | 45.1 | 27.6 | Intended product: 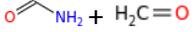 |
| | 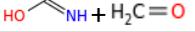 | | | |
| 18 | O⧸⧹NH$_2$ + H$_2$C=O | 47.4 | -60.8 | |
| 19 | HO⧸⧹NH + H$_2$C=O | 48.1 | -49.3 | |
| 20 | O⧸⧹NH⧸⧹OH | 45.8 | -72.2 | |
| 21 | 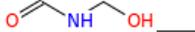 | 66.6 | 12.4 | |

[a] in kcal/mol; geometry optimization from Gaussian 03 using M062X level of theory, energy from Gaussian 03 using CCSD(T) level of theory. Basis set is 6-311++G* in both calculations.

[b] in kcal/mol; thermochemistry from Gaussian 03 using CBS-QB3 level of theory.

**Table 4.** Automatically identified unimolecular reactions of 1,2-dioxolan-3-ol (OC1OOCC1) and 1,2,4-dioxazolidin-3-ol (OC1OOCN1). For mechanistic details see Figure 4.

| | Product(s) | ΔE$^{\#a}$ | ΔH$_r^0$(298 K)$^b$ | Comments |
|---|---|---|---|---|
| | 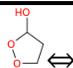 ⇔ | | | |
| 22 | 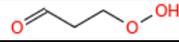 | 45.1 | 10.4 | |
| 23 | 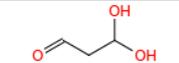 | 77.0 | -52.2 | |
| 24 | 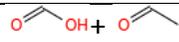 | 38.2 | -54.7 | |
| 25 | 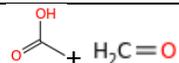 | 41.3 | -53.2 | |
| | 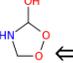 ⇔ | | | |
| 26 | 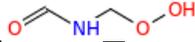 | 17.6 | -10.4 | |
| 27 | 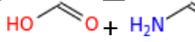 | 27.1 | -79.8 | |
| 28 | 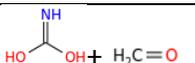 | 29.0 | -46.9 | |
| 29 | 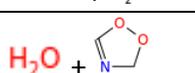 | 38.6 | 2.8 | |
| 30 | 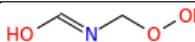 | 40.2 | -1.5 | |
| 31 | 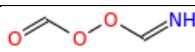 +H$_2$ | 52.2 | 12.1 | |
| 32 | 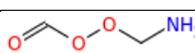 | 54.9 | -9.6 | |
| 33 | 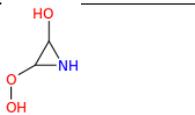 | 64.1 | 19.5 | |
| 34 | 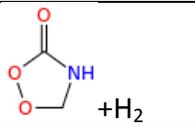 +H$_2$ | 68.7 | -2.3 | |
| 35 | 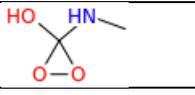 | 89.0 | 0.2 | |
| 36 | 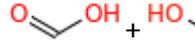 | 33.8 | -64.2 | |

$^a$ in kcal/mol; geometry optimization from Gaussian 03 using M062X level of theory, energy from Gaussian 03 using CCSD(T) level of theory. Basis set is 6-311++G* in both calculations.

$^b$ in kcal/mol; thermochemistry from Gaussian 03 using CBS-QB3 level of theory.

**Table 5.** Automatically identified unimolecular reactions of penta-1,4-diene (C=CCC=C). For mechanistic details see Figure 5.

| | Product(s) | $\Delta E^{\#a}$ | $\Delta H_r^0(298\ K)^b$ | Comments |
|---|---|---|---|---|
| | 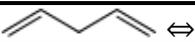 ⇔ | | | |
| 37 | 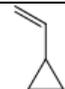 | 63.0 | 5.2 | |
| 38 | 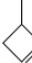 | 95.0 | 5.6 | |
| 39 |  | 104.3 | -17.4 | Unexpected H-shift |
| 40 | 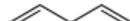 | 86.6 | 0.0 | Framework rearrangement |
| 41 | 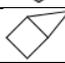 | 66.2 | 11.3 | |
| 42 | 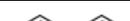 | 68.4 | -7.1 | Intended product: 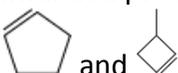 and 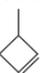 |
| 43 | 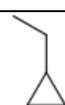 | 101.8 | 29.5 | Intended product: 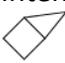 |
| 44 | 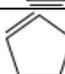 | 61.6 | -17.4 | 1,2 H-shift |

[a] in kcal/mol; geometry optimization from Gaussian 03 using M062X level of theory, energy from Gaussian 03 using CCSD(T) level of theory. Basis set is 6-311++G* in both calculations.

[b] in kcal/mol; thermochemistry from Gaussian 03 using CBS-QB3 level of theory.

# Supporting Information

# Automated Discovery of Elementary Chemical Reaction Steps Using Freezing String and Berny Optimization Methods

Yury V. Suleimanov[1,2,*] and William H. Green[1]

[1]Department of Chemical Engineering, Massachusetts Institute of Technology, Cambridge, Massachusetts 02139, Unites States

[2]Computation-based Science and Technology Research Center, Cyprus Institute, 20 Kavafi Street, Nicosia 2121, Cyprus

* Email: ysuleyma@mit.edu

**Cartesian coordinates (Å) of all the automatically detected saddle points.**

Reaction (1)

| | | | |
|---|---|---|---|
| O |  0.00000000 |  0.00000000 | 0.00000000 |
| C |  0.00000000 |  0.00000000 | 1.29965600 |
| C |  1.32028500 |  0.00000000 | 2.04187100 |
| C |  1.93608600 |  1.34729500 | 1.67961100 |
| O |  0.89286200 |  2.30395700 | 1.87310900 |
| O | -0.26814300 |  1.69832000 | 1.35079400 |
| H | -0.86796700 | -0.40626100 | 1.83054700 |
| H |  1.16540900 | -0.07805900 | 3.11885400 |
| H |  1.92976200 | -0.83135300 | 1.68637300 |
| H |  2.26547800 |  1.35715200 | 0.63670600 |
| H |  2.74538800 |  1.66009800 | 2.33743300 |
| H | -0.12528400 |  1.28884400 | 0.25957400 |

Reaction (2)

| | | | |
|---|---|---|---|
| O |  0.00000000 |  0.00000000 | 0.00000000 |
| C |  0.00000000 |  0.00000000 | 1.19802500 |
| C |  1.25213000 |  0.00000000 | 2.04909200 |
| C |  1.15164700 | -1.07543200 | 3.12301500 |
| O |  0.27113200 | -0.88219500 | 4.06456900 |
| O |  1.63641900 | -0.47830900 | 5.22537600 |
| H | -0.94873400 | -0.00854900 | 1.77328400 |
| H |  2.12589800 | -0.13806900 | 1.41154300 |
| H |  1.31553300 |  0.97105600 | 2.55194400 |
| H | 2.11935700 | -0.88673800 | 3.92625900 |
| H | 1.28101000 | -2.12357500 | 2.82280900 |
| H | 1.72849600 | -0.24489800 | 6.17058400 |

Reaction (3)

| | | | |
|---|---:|---:|---:|
| O | -1.05363700 | 1.37481200 | 0.18535300 |
| C | -0.17753600 | 0.82633100 | -0.45078300 |
| C | 1.41220000 | 0.58482100 | 0.43775500 |
| C | 1.41786400 | -0.74259300 | -0.00915600 |
| O | -0.42022100 | -0.71215800 | -0.67247300 |
| O | -1.12261800 | -1.19248300 | 0.45390900 |
| H | 0.16983400 | 1.13642000 | -1.44507800 |
| H | 2.07631900 | 1.29713800 | -0.03734700 |
| H | 1.19980200 | 0.76892700 | 1.48331800 |
| H | 1.17599000 | -1.56948900 | 0.64815600 |
| H | 1.86941000 | -1.00855600 | -0.95924900 |
| H | -1.63471700 | -0.39716500 | 0.70898600 |

Reaction (4)

| | | | |
|---|---:|---:|---:|
| O | 1.55056400 | 0.26109400 | 0.84895000 |
| C | 0.99666100 | 0.12831300 | -0.25900500 |
| C | -0.22743000 | 1.07155400 | -0.63498400 |
| C | -1.26303000 | 0.51273900 | 0.39584200 |
| O | -1.35210000 | -0.81048300 | 0.28048900 |
| O | 0.35898100 | -1.15083100 | -0.53929400 |
| H | 1.60393100 | 0.23178500 | -1.20759300 |
| H | 0.01527800 | 2.11858000 | -0.45793500 |
| H | -0.56094400 | 0.89218400 | -1.65725500 |
| H | -2.19680800 | 0.96472600 | 0.00704100 |
| H | -0.97159700 | 0.86052600 | 1.39274000 |
| H | 0.61336800 | -1.74167800 | 0.19072400 |

Reaction (5)

| | | | |
|---|---:|---:|---:|
| O | 0.00000000 | 0.00000000 | 0.00000000 |
| C | 0.00000000 | 0.00000000 | 1.19694200 |
| C | 0.86817800 | 0.00000000 | 2.27933200 |
| C | 2.34046100 | -0.19779400 | 2.11954900 |
| O | 2.62200400 | -1.51769700 | 2.56843100 |
| O | 4.03786100 | -1.66483600 | 2.52724100 |
| H | -1.14372100 | 0.30662900 | 1.75859100 |
| H | -0.44690200 | 0.99254400 | 2.05996800 |
| H | 0.42526300 | -0.12205600 | 3.26051300 |
| H | 2.89168100 | 0.51351500 | 2.74253200 |
| H | 2.63950800 | -0.07910700 | 1.07459100 |
| H | 4.13703800 | -2.38361500 | 1.89114000 |

Reaction (6)

| | | | |
|---|---:|---:|---:|
| O | 0.00000000 | 0.00000000 | 0.00000000 |
| C | 0.00000000 | 0.00000000 | 1.26651400 |
| C | 1.33263800 | 0.00000000 | 1.75258200 |
| C | 1.73044400 | -0.58736600 | 3.06795000 |
| O | 1.47219700 | 0.39476700 | 4.07065100 |
| O | 1.97846300 | -0.13701500 | 5.29687900 |
| H | -0.94895500 | -0.03325900 | 1.80792100 |
| H | 1.23222800 | -0.21251300 | 0.28310200 |
| H | 1.81910700 | 0.94899500 | 1.51301100 |
| H | 2.79657700 | -0.82811900 | 3.07935400 |
| H | 1.16610000 | -1.49681800 | 3.30002800 |
| H | 1.21756800 | -0.02514100 | 5.87803300 |

Reaction (7)

| | | | |
|---|---|---|---|
| O | 0.00000000 | 0.00000000 | 0.00000000 |
| C | 0.00000000 | 0.00000000 | 1.29404900 |
| C | 2.27814000 | 0.00000000 | 1.70212400 |
| O | 1.99061700 | 1.37748500 | 1.90721000 |
| O | 0.69999000 | 1.53773700 | 1.36334800 |
| H | -0.94642000 | 0.07148600 | 1.83542100 |
| H | 2.57764400 | -0.15926900 | 0.65852700 |
| H | 3.08242600 | -0.25251900 | 2.39074600 |
| H | 0.61776200 | 1.15252400 | 0.26532600 |
| N | 1.03145500 | -0.66068100 | 2.05928000 |
| H | 1.04589300 | -1.64712000 | 1.81632400 |

Reaction (8)

| | | | |
|---|---|---|---|
| O | 0.00000000 | 0.00000000 | 0.00000000 |
| O | 0.00000000 | 0.00000000 | 1.83939200 |
| C | 1.29372500 | 0.00000000 | 1.82503100 |
| N | 1.95364300 | 1.22333100 | 2.16428000 |
| C | 1.48952900 | 2.08085000 | 3.12737100 |
| O | 2.00993300 | 3.13185500 | 3.39738500 |
| H | -0.66655800 | 0.01194100 | -0.71477500 |
| H | 1.33382000 | -0.13336200 | 0.56831800 |
| H | 1.75939800 | -0.92200600 | 2.20032900 |
| H | 2.78621500 | 1.50824000 | 1.66711500 |
| H | 0.58814100 | 1.69771100 | 3.63232700 |

Reaction (9)

| | | | |
|---|---|---|---|
| O | 0.00000000 | 0.00000000 | 0.00000000 |
| O | 0.00000000 | 0.00000000 | 1.68332100 |
| C | 1.22924300 | 0.00000000 | 2.05072300 |
| N | 1.88915900 | -1.27045800 | 1.64286100 |
| C | 2.47616600 | -1.97481900 | 2.54488900 |
| O | 2.51989200 | -1.54928400 | 3.76716300 |
| H | -0.94962400 | 0.01937200 | -0.16307400 |
| H | 1.93244100 | -0.72762600 | 3.74993700 |
| H | 1.86756600 | 0.83833200 | 1.76005800 |
| H | 1.72918300 | -1.61335200 | 0.69728100 |
| H | 2.93989700 | -2.93242900 | 2.33672100 |

Reaction (10)

| | | | |
|---|---|---|---|
| O | 0.00000000 | 0.00000000 | 0.00000000 |
| O | 0.00000000 | 0.00000000 | 1.40996200 |
| C | 1.97522200 | 0.00000000 | 1.96166500 |
| N | 1.91346300 | -1.27904800 | 1.71956400 |
| C | 0.36220700 | -1.60089900 | 1.83567500 |
| O | -0.06104100 | -2.40069700 | 1.02595200 |
| H | -0.37067500 | -0.88380400 | -0.19505500 |
| H | 1.58326000 | 0.37164400 | 2.90181700 |
| H | 2.59723600 | 0.66873400 | 1.37238600 |
| H | 2.28785200 | -1.60719200 | 0.83347600 |
| H | 0.09324400 | -1.55072000 | 2.89883800 |

Reaction (11)

| | | | |
|---|---:|---:|---:|
| O | 0.00000000 | 0.00000000 | 0.00000000 |
| O | 0.00000000 | 0.00000000 | 1.97894900 |
| C | 1.32954800 | 0.00000000 | 1.93619100 |
| N | 1.98131400 | -1.08377000 | 1.16306700 |
| C | 1.92876600 | -1.37499600 | -0.18324200 |
| O | 2.54201300 | -2.16988600 | -0.79803600 |
| H | -0.69153600 | 0.39071400 | -0.56694600 |
| H | 1.68091000 | -0.20896200 | 2.96695100 |
| H | 1.75766700 | 0.96022100 | 1.61897500 |
| H | 2.73758300 | -1.58335200 | 1.62447600 |
| H | 0.82797300 | -0.59108700 | -0.57350600 |

Reaction (12)

| | | | |
|---|---:|---:|---:|
| O | 0.00000000 | 0.00000000 | 0.00000000 |
| O | 0.00000000 | 0.00000000 | 1.42007200 |
| C | 1.84164400 | 0.00000000 | -0.56082400 |
| N | 1.87421200 | 0.96783000 | -1.45094800 |
| C | 0.73066600 | 2.20379100 | -0.94449000 |
| O | 0.82152800 | 3.23352900 | -1.46396500 |
| H | 0.09348200 | 1.20109300 | -0.27024900 |
| H | 1.95620600 | -1.03966000 | -0.85492100 |
| H | 2.11251800 | 0.23909300 | 0.46473700 |
| H | 1.76710800 | 0.67520500 | -2.41807800 |
| H | -0.67549000 | -0.65983900 | 1.61265500 |

Reaction (13)

| | | | |
|---|---:|---:|---:|
| O | 0.00000000 | 0.00000000 | 0.00000000 |
| O | 0.00000000 | 0.00000000 | 1.42541500 |
| C | 1.37764400 | 0.00000000 | 1.82777500 |
| N | 1.42549100 | -0.04972600 | 3.24164500 |
| C | 1.04776700 | -1.81328500 | 3.81275200 |
| O | 0.01786300 | -2.14312200 | 4.18470800 |
| H | -0.49014800 | 0.80764300 | -0.19213200 |
| H | 1.86559100 | 0.88220100 | 1.39801400 |
| H | 1.84921800 | -0.90280900 | 1.41018800 |
| H | 1.11016000 | 0.79468500 | 3.71235100 |
| H | 2.08421700 | -1.26524700 | 3.64042500 |

Reaction (14)

| | | | |
|---|---:|---:|---:|
| O | 0.00000000 | 0.00000000 | 0.00000000 |
| O | 0.00000000 | 0.00000000 | 1.44578900 |
| C | 1.42236200 | 0.00000000 | 1.83338700 |
| N | 2.14097100 | -1.00855900 | 1.17710500 |
| C | 2.40751800 | -0.85606100 | -0.27252800 |
| O | 3.27826900 | -1.54636600 | -0.72368900 |
| H | 0.50110900 | -0.85057700 | -0.29629400 |
| H | 1.36549900 | -0.14881700 | 2.91091300 |
| H | 1.76970300 | 1.01465000 | 1.60194900 |
| H | 2.68104700 | -1.67444900 | 1.72214300 |
| H | 0.80180200 | 0.55094800 | -0.32101500 |

Reaction (15)

| | | | |
|---|---:|---:|---:|
| C | 0.00000000 | 0.00000000 | 0.00000000 |
| O | 0.00000000 | 0.00000000 | 1.30944600 |
| O | 2.23151100 | 0.00000000 | 1.24181500 |
| C | 1.89888800 | 1.02568600 | 0.56146300 |
| O | 1.09495900 | 0.80181300 | -0.52492800 |
| H | 0.07694700 | -1.01597500 | -0.43202200 |
| H | -0.90024000 | 0.48006000 | -0.42201600 |
| H | 1.08666700 | 1.50697900 | 1.31519200 |
| H | 2.62795100 | 1.83887200 | 0.43285600 |

Reaction (16)

| | | | |
|---|---:|---:|---:|
| C | 0.00000000 | 0.00000000 | 0.00000000 |
| O | 0.00000000 | 0.00000000 | 1.28826800 |
| O | 1.90574300 | 0.00000000 | 1.41698100 |
| C | 1.98741100 | 1.02651700 | 0.58538200 |
| O | 1.33306200 | 0.86898000 | -0.51868600 |
| H | 0.17838100 | -0.94388600 | -0.52524900 |
| H | -0.75458600 | 0.64492500 | -0.46167300 |
| H | 2.08208300 | 2.03319400 | 1.01034800 |
| H | 3.06745700 | 0.44306600 | 0.76831400 |

Reaction (17)

| | | | |
|---|---:|---:|---:|
| C | 0.00000000 | 0.00000000 | 0.00000000 |
| O | 0.00000000 | 0.00000000 | 1.24763700 |
| O | 1.24562200 | 0.00000000 | 1.80463800 |
| C | 1.82352800 | 1.87495400 | 0.99151900 |
| O | 1.09784600 | 1.97903600 | 0.00332500 |
| H | 0.90044300 | -0.26093700 | -0.54054600 |
| H | -0.97055200 | 0.13512400 | -0.46416800 |
| H | 1.56311400 | 2.36878100 | 1.93605300 |
| H | 2.84870700 | 1.49037800 | 0.90864500 |

Reaction (18)

| | | | |
|---|---:|---:|---:|
| C | 0.00000000 | 0.00000000 | 0.00000000 |
| O | 0.00000000 | 0.00000000 | 1.27139600 |
| O | 2.12009300 | 0.00000000 | 1.22197700 |
| C | 2.39626400 | -0.31603700 | 0.03346600 |
| N | 0.82070000 | -0.95659500 | -0.66259200 |
| H | 0.79811000 | -1.78694700 | -0.06115400 |
| H | -0.05348100 | 0.95044500 | -0.54619100 |
| H | 2.90312300 | -1.27739500 | -0.14372300 |
| H | 2.60036600 | 0.46986600 | -0.70381700 |
| H | -0.82127900 | -0.71301400 | -0.42422600 |

Reaction (19)

| | | | |
|---|---|---|---|
| C | 0.00000000 | 0.00000000 | 0.00000000 |
| O | 0.00000000 | 0.00000000 | 1.32707700 |
| O | 1.93710300 | 0.00000000 | 1.39516900 |
| C | 2.03454000 | 1.00837600 | 0.55517700 |
| N | 0.83371900 | 0.94336000 | -0.48878900 |
| H | -0.16924200 | -0.94258900 | -0.53208400 |
| H | -1.07754000 | 0.30932700 | 0.57665300 |
| H | 1.98221000 | 2.01345900 | 0.99133100 |
| H | 2.89940100 | 0.91008400 | -0.11704400 |
| H | 1.08864600 | 0.81096500 | -1.46097700 |

Reaction (20)

| | | | |
|---|---|---|---|
| C | 0.00000000 | 0.00000000 | 0.00000000 |
| O | 0.00000000 | 0.00000000 | 1.31079800 |
| O | 2.31752900 | 0.00000000 | 1.30220100 |
| C | 2.01664300 | -1.01112900 | 0.57334300 |
| N | 1.18225800 | -0.73476200 | -0.52785600 |
| H | -0.87939500 | -0.53076400 | -0.41555100 |
| H | -0.02157100 | 1.01797400 | -0.43684700 |
| H | 2.77380000 | -1.80196500 | 0.44806200 |
| H | 1.20422900 | -1.53875800 | 1.26900700 |
| H | 1.01853400 | -1.48150500 | -1.19242600 |

Reaction (21)

| | | | |
|---|---|---|---|
| C | 0.94750700 | -0.72695500 | 0.07368700 |
| O | 0.93297200 | 0.79938700 | 0.16182800 |
| O | -0.39227600 | 1.11423500 | -0.24311900 |
| C | -1.15239400 | -0.25323500 | 0.36245200 |
| N | -0.34615900 | -1.05003000 | -0.41069500 |
| H | 1.73827400 | -1.00536700 | -0.62047200 |
| H | 1.16699000 | -1.05017300 | 1.09614900 |
| H | -0.98983800 | -0.14525800 | 1.43943400 |
| H | -2.18342800 | -0.13889200 | 0.03511600 |
| H | -0.40512400 | 0.26206800 | -1.04186900 |

Reaction (22)

| | | | |
|---|---|---|---|
| C | -0.13476700 | 1.24233700 | -0.36294600 |
| C | 0.99772000 | 0.23964500 | -0.43938700 |
| O | 1.64271900 | -0.07660400 | 0.64362300 |
| O | 0.02836100 | -1.17783400 | -0.33680300 |
| O | -1.29460500 | -0.77776500 | -0.05882700 |
| C | -1.18289000 | 0.53622600 | 0.49057100 |
| H | 0.22687900 | 2.15384300 | 0.11390900 |
| H | -0.52937300 | 1.46560700 | -1.35512300 |
| H | 1.49965100 | 0.13025000 | -1.40704200 |
| H | 0.75250900 | -1.04950000 | 0.57879800 |
| H | -0.86169600 | 0.48038700 | 1.53458200 |
| H | -2.18015000 | 0.96779000 | 0.42150300 |

Reaction (23)

| | | | |
|---|---|---|---|
| C | 0.15052300 | 1.16823600 | 0.19522100 |
| C | -0.76527500 | -0.08046600 | 0.41359200 |
| O | -1.98153800 | 0.17346100 | -0.23869300 |
| O | -0.14996200 | -1.18434400 | -0.18520700 |
| O | 1.66762100 | -0.54894100 | 0.32964800 |
| C | 1.33033100 | 0.45724100 | -0.42621700 |
| H | -0.33148900 | 1.89283800 | -0.45982600 |
| H | 0.41281200 | 1.63044300 | 1.14738400 |
| H | -0.92176300 | -0.30782500 | 1.47381200 |
| H | -2.50315200 | -0.63555000 | -0.21260100 |
| H | 0.65701700 | -0.23158700 | -1.20265100 |
| H | 2.10412800 | 0.86021200 | -1.08768000 |

Reaction (24)

| | | | |
|---|---|---|---|
| C | 0.43827600 | 1.05860900 | 0.62288300 |
| C | -0.97581600 | -0.04699100 | 0.41959000 |
| O | -1.64390400 | 0.37439600 | -0.71591000 |
| O | -0.50380100 | -1.19583200 | 0.48022800 |
| O | 1.48174000 | -0.83904900 | -0.27471200 |
| C | 1.25925000 | 0.40277700 | -0.46384000 |
| H | 0.16146600 | 2.10218300 | 0.47293300 |
| H | 0.76930600 | 0.80777300 | 1.62864100 |
| H | -1.43681000 | 0.39614800 | 1.32256100 |
| H | -1.58644500 | -0.34977200 | -1.35091800 |
| H | 1.01170600 | 0.77348400 | -1.46829900 |
| H | 2.07822800 | 1.06769300 | 0.00643600 |

Reaction (25)

| | | | |
|---|---|---|---|
| C | 1.43231200 | 0.41203500 | 0.43079600 |
| C | 0.13879700 | 1.08299200 | -0.64824300 |
| H | -0.11524900 | 2.10243600 | -0.36021900 |
| H | 0.54475800 | 0.96665500 | -1.64902700 |
| C | -0.86178800 | 0.00339400 | -0.32298600 |
| O | -1.53801200 | 0.26450500 | 0.85599000 |
| O | 1.45767100 | -0.83482300 | 0.26290200 |
| O | -0.52355600 | -1.16121100 | -0.62637600 |
| H | -1.76057300 | -0.59501900 | 1.23149400 |
| H | 1.12292200 | 0.81676400 | 1.40618700 |
| H | 2.18625300 | 1.01845400 | -0.11251000 |
| H | -1.40285800 | 0.55240800 | -1.21344700 |

Reaction (26)

| | | | |
|---|---|---|---|
| C | 0.00000000 | 0.00000000 | 0.00000000 |
| N | 0.00000000 | 0.00000000 | 1.46624200 |
| C | 1.23964700 | 0.00000000 | 1.97632800 |
| O | 1.90100800 | 1.14650500 | 0.33041800 |
| O | 0.69771700 | 1.14547200 | -0.39233500 |
| O | 2.17361900 | -0.78938100 | 1.53039200 |
| H | -1.01415500 | 0.06992500 | -0.38611700 |
| H | 0.49296600 | -0.91417700 | -0.33986200 |
| H | -0.60840500 | 0.69913100 | 1.87673900 |
| H | 1.37301200 | 0.48327300 | 2.94363900 |
| H | 2.36125800 | -0.17261800 | 0.65974600 |

Reaction (27)

| | | | |
|---|---|---|---|
| C | 0.00000000 | 0.00000000 | 0.00000000 |
| N | 0.00000000 | 0.00000000 | 1.42365800 |
| C | 1.81754700 | 0.00000000 | 1.60826600 |
| O | 2.19696700 | 0.83852200 | 0.76609800 |
| O | 0.57845900 | 1.01909000 | -0.51192200 |
| O | 2.16299700 | -1.32848100 | 1.44506000 |
| H | -1.10950900 | 0.29645600 | 0.12620600 |
| H | 0.10174400 | -0.97570400 | -0.50215300 |
| H | -0.31814700 | -0.91980400 | 1.75042800 |
| H | 1.75128400 | 0.26747500 | 2.66764400 |
| H | 2.66767800 | -1.38807900 | 0.62493600 |

Reaction (28)

| | | | |
|---|---|---|---|
| C | 0.00000000 | 0.00000000 | 0.00000000 |
| N | 0.00000000 | 0.00000000 | 1.59869200 |
| C | 1.31046100 | 0.00000000 | 1.92349900 |
| O | 2.03865800 | 0.69511800 | 1.07742500 |
| O | 1.22782200 | -0.18388000 | -0.44708200 |
| O | 1.81646800 | -1.10285800 | 2.52656300 |
| H | -0.47669300 | 0.94650400 | -0.28038500 |
| H | -0.65229100 | -0.86076900 | -0.20784600 |
| H | -0.50186300 | -0.76823700 | 2.03192400 |
| H | 1.81663200 | 1.08724800 | 2.36815500 |
| H | 2.77787600 | -1.08934000 | 2.44166900 |

Reaction (29)

| | | | |
|---|---|---|---|
| C | 0.00000000 | 0.00000000 | 0.00000000 |
| N | 0.00000000 | 0.00000000 | 1.44848900 |
| C | 1.32108400 | 0.00000000 | 1.75048200 |
| O | 2.12470600 | -0.02335900 | 0.69013800 |
| O | 1.29341100 | -0.48673700 | -0.38459400 |
| O | 1.11685900 | -1.68189500 | 2.44608800 |
| H | -0.14174800 | 1.01068200 | -0.39886000 |
| H | -0.72310500 | -0.69576700 | -0.42378200 |
| H | 0.08259700 | -1.15283000 | 2.04803700 |
| H | 1.78603900 | 0.47533500 | 2.60723200 |
| H | 1.28195300 | -1.88423400 | 3.37532200 |

Reaction (30)

| | | | |
|---|---|---|---|
| C | 0.00000000 | 0.00000000 | 0.00000000 |
| N | 0.00000000 | 0.00000000 | 1.54970300 |
| C | 1.13093900 | 0.00000000 | 2.14658800 |
| O | 0.31533500 | -2.15555800 | 0.31316000 |
| O | 0.81932000 | -0.97506600 | -0.39174700 |
| O | 1.36344800 | -0.42847000 | 3.35516900 |
| H | 0.35417700 | 0.98171700 | -0.32934800 |
| H | -1.05609500 | -0.16682500 | -0.23930000 |
| H | -0.75152700 | -0.55772200 | 1.94611700 |
| H | 2.00328200 | 0.38964400 | 1.63186500 |
| H | 0.62027500 | -0.93355900 | 3.71384400 |

Reaction (31)

| | | | |
|---|---|---|---|
| C | 0.00000000 | 0.00000000 | 0.00000000 |
| N | 0.00000000 | 0.00000000 | 1.32615300 |
| C | 1.52955300 | 0.00000000 | 1.60061900 |
| O | 1.82204200 | 1.06328200 | 0.62149000 |
| O | 0.95701400 | 0.76990500 | -0.48389000 |
| O | 2.04786600 | -1.14250400 | 1.34640500 |
| H | -0.84510600 | -0.19091600 | -0.64538800 |
| H | 0.64009600 | -1.57170600 | -0.34865500 |
| H | -0.54070600 | -0.72578000 | 1.78114900 |
| H | 1.73223300 | 0.47196500 | 2.56578100 |
| H | 1.28564900 | -1.57115600 | 0.28248700 |

Reaction (32)

| | | | |
|---|---|---|---|
| C | 0.00000000 | 0.00000000 | 0.00000000 |
| N | 0.00000000 | 0.00000000 | 1.44741600 |
| C | 1.50989800 | 0.00000000 | 1.74018600 |
| O | 1.86157700 | -1.11867500 | 0.92735200 |
| O | 1.22466700 | -0.74480200 | -0.31031600 |
| O | 1.48703200 | -0.01345100 | 3.04321100 |
| H | 0.09609400 | 1.00567300 | -0.40521500 |
| H | -0.85287300 | -0.52893000 | -0.42187400 |
| H | -0.27743300 | -0.93013000 | 1.76873900 |
| H | 1.93246200 | 0.89264700 | 1.24648200 |
| H | 0.20769500 | 0.50684500 | 2.53195400 |

Reaction (33)

| | | | |
|---|---|---|---|
| C | 0.00000000 | 0.00000000 | 0.00000000 |
| N | 0.00000000 | 0.00000000 | 1.52019300 |
| C | 1.12659600 | 0.00000000 | 2.12436800 |
| O | 1.67304700 | -1.22320100 | -0.39910100 |
| O | 0.21646100 | -1.31654800 | -0.35784700 |
| O | 1.20111300 | -0.33582100 | 3.39483000 |
| H | 1.31235200 | 0.00345800 | -0.38800100 |
| H | -0.97234200 | 0.37232100 | -0.32145100 |
| H | -0.80254400 | -0.34003000 | 2.04942500 |
| H | 2.01794700 | 0.27702100 | 1.57344800 |
| H | 2.11250100 | -0.31852100 | 3.70545600 |

Reaction (34)

| | | | |
|---|---|---|---|
| C | 0.00000000 | 0.00000000 | 0.00000000 |
| N | 0.00000000 | 0.00000000 | 1.45350300 |
| C | 1.32246800 | 0.00000000 | 1.85964300 |
| O | 2.10250300 | 0.03589000 | 0.72225000 |
| O | 1.25363800 | 0.59426600 | -0.27873100 |
| O | 1.75717400 | -0.67157600 | 2.87893000 |
| H | -0.75432600 | 0.65580600 | -0.43287800 |
| H | -0.05762700 | -1.01399300 | -0.41186900 |
| H | -0.56738000 | -0.66399800 | 1.96331900 |
| H | 1.57829600 | 1.36645600 | 2.38788800 |
| H | 1.86651400 | 0.62482800 | 3.01227000 |

Reaction (35)

| | | | |
|---|---:|---:|---:|
| C | 0.00000000 | 0.00000000 | 0.00000000 |
| N | 0.00000000 | 0.00000000 | 1.42792500 |
| C | 1.42456600 | 0.00000000 | 1.40713300 |
| O | 2.02506500 | -1.20378900 | 1.23950000 |
| O | 1.23290800 | -1.88196000 | 0.33182500 |
| O | 2.07276100 | 0.82215800 | 2.24888800 |
| H | -0.08298200 | 0.99304300 | -0.43647400 |
| H | -0.47266300 | -0.80469600 | -0.54522000 |
| H | -0.34558300 | -0.88123500 | 1.80103900 |
| H | 1.47415100 | 0.31578000 | 0.16794100 |
| H | 1.41270800 | 1.30493700 | 2.75831600 |

Reaction (36)

| | | | |
|---|---:|---:|---:|
| C | 0.00000000 | 0.00000000 | 0.00000000 |
| N | 0.00000000 | 0.00000000 | 1.34879800 |
| C | 1.58806300 | 0.00000000 | 1.76564000 |
| O | 2.28588500 | 0.04652700 | 0.67329900 |
| O | 0.96801300 | -0.77765300 | -0.47135800 |
| O | 1.79987000 | -1.07108900 | 2.60607500 |
| H | -0.28872900 | 0.88447200 | -0.57966900 |
| H | -0.40030900 | -1.00191400 | -0.60250800 |
| H | -0.54797900 | 0.74081500 | 1.77416100 |
| H | 1.62071300 | 0.92346200 | 2.35814800 |
| H | 2.08126000 | -1.80473200 | 2.04764200 |

Reaction (37)

| | | | |
|---|---:|---:|---:|
| C | 0.00000000 | 0.00000000 | 0.00000000 |
| C | 0.00000000 | 0.00000000 | 1.43289800 |
| C | 1.28635800 | 0.00000000 | 2.15717600 |
| C | 1.44119200 | 0.61226600 | 3.42801600 |
| C | 2.62225200 | 0.78399600 | 4.06339300 |
| H | 0.76221200 | 0.56039000 | -0.52619900 |
| H | -0.69506100 | -0.59169600 | -0.58009200 |
| H | -0.19332300 | 1.09631600 | 1.62053500 |
| H | -0.82042700 | -0.53102400 | 1.92291100 |
| H | 2.17996400 | -0.28716800 | 1.60918800 |
| H | 0.52562300 | 0.90311100 | 3.94393600 |
| H | 2.66879400 | 1.14720700 | 5.08139100 |
| H | 3.55968600 | 0.54677800 | 3.57250400 |

Reaction (38)

| | | | |
|---|---|---|---|
| C | 0.00000000 | 0.00000000 | 0.00000000 |
| C | 0.00000000 | 0.00000000 | 1.54365500 |
| C | 1.44770600 | 0.00000000 | 2.05888000 |
| C | -0.16556300 | -1.46963100 | 1.49375600 |
| C | 0.49291600 | -1.45926100 | 0.23254800 |
| H | 0.68844400 | 0.68512000 | -0.49707000 |
| H | -1.00834500 | 0.04899100 | -0.41315100 |
| H | -0.63972800 | 0.66426000 | 2.12400900 |
| H | 1.84017800 | -0.93731400 | 2.45005800 |
| H | 1.72024000 | 0.84038400 | 2.70678100 |
| H | -0.34281500 | -2.21692500 | 2.25805600 |
| H | 0.62525900 | -2.27184600 | -0.47493500 |
| H | 1.59470000 | -0.99555300 | 0.65488900 |

Reaction (39)

| | | | |
|---|---|---|---|
| C | 0.00000000 | 0.00000000 | 0.00000000 |
| C | 0.00000000 | 0.00000000 | 1.43233500 |
| C | 1.19513600 | 0.00000000 | 2.27938100 |
| C | 2.50587200 | 0.16768400 | 1.82539700 |
| C | 2.96575600 | 0.28758300 | 0.54153800 |
| H | 0.68925100 | -0.60733200 | -0.56146400 |
| H | -0.86737500 | 0.37112600 | -0.52948100 |
| H | -0.15553400 | -1.10734600 | 1.64227600 |
| H | -0.84949400 | 0.50301300 | 1.89796400 |
| H | 1.02017100 | -0.15093300 | 3.33779800 |
| H | 3.23618600 | 0.32072700 | 2.61786800 |
| H | 3.96857100 | 0.66479600 | 0.37905700 |
| H | 2.42424300 | -0.02566300 | -0.33154800 |

Reaction (40)

| | | | |
|---|---|---|---|
| C | 0.00000000 | 0.00000000 | 0.00000000 |
| C | 0.00000000 | 0.00000000 | 1.40423200 |
| C | 1.60335200 | 0.00000000 | 2.07058300 |
| C | 0.94211900 | -0.99971600 | 2.86787100 |
| C | -0.09525600 | -1.60036000 | 2.07075200 |
| H | -0.54755300 | -0.76228300 | -0.53956800 |
| H | 0.79374200 | 0.50069400 | -0.53969600 |
| H | -0.59023900 | 0.62646100 | 2.06082800 |
| H | 2.23771000 | -0.29102600 | 1.24548400 |
| H | 1.96688100 | 0.88360000 | 2.59246300 |
| H | 0.60572200 | -0.64264900 | 3.83867700 |
| H | -0.99883200 | -1.91092000 | 2.59263100 |
| H | 0.15740300 | -2.25071900 | 1.24542600 |

Reaction (41)

| | | | |
|---|---|---|---|
| C | 0.00000000 | 0.00000000 | 0.00000000 |
| C | 0.00000000 | 0.00000000 | 1.42881500 |
| C | 1.28563400 | 0.00000000 | 2.19351600 |
| C | 1.89448300 | -0.86399100 | 1.15201700 |
| C | 1.95198800 | -0.19655400 | -0.07970200 |
| H | 0.02867800 | -0.91045300 | -0.58312100 |
| H | -0.73290900 | 0.68284000 | -0.45915100 |
| H | -0.94336900 | 0.11844100 | 1.96395100 |
| H | 1.75054100 | 0.98869800 | 2.29970700 |
| H | 1.17879800 | -0.45562000 | 3.17814400 |
| H | 1.89207400 | -1.94166900 | 1.24524000 |
| H | 2.20753300 | -0.74001300 | -0.98297200 |
| H | 2.27376900 | 0.84276200 | -0.07832700 |

Reaction (42)

| | | | |
|---|---|---|---|
| C | 0.00000000 | 0.00000000 | 0.00000000 |
| C | 0.00000000 | 0.00000000 | 1.35334300 |
| C | 1.11192300 | 0.00000000 | 2.24343500 |
| C | 2.48801200 | -0.22919300 | 1.90790800 |
| C | 3.11918500 | -0.51845200 | 0.66382000 |
| H | -0.94082300 | 0.02152800 | -0.53525500 |
| H | 0.89526900 | 0.03560700 | -0.59959300 |
| H | -0.96984300 | 0.01549900 | 1.84461500 |
| H | 3.17523200 | -0.08622900 | 2.73850600 |
| H | 0.91110000 | 0.08302800 | 3.30331800 |
| H | 2.19049900 | -1.35726500 | 1.91433200 |
| H | 4.17533000 | -0.74736200 | 0.67959100 |
| H | 2.58025900 | -0.68776700 | -0.25159000 |

Reaction (43)

| | | | |
|---|---|---|---|
| C | 0.00000000 | 0.00000000 | 0.00000000 |
| C | 0.00000000 | 0.00000000 | 1.45326700 |
| C | 1.28698600 | 0.00000000 | 2.19184100 |
| C | 2.54368900 | -0.38153200 | 1.60047200 |
| C | 2.54910600 | 0.67128500 | 0.77508000 |
| H | -0.94034800 | 0.25530400 | -0.47515600 |
| H | 0.57300400 | -0.77183000 | -0.50671800 |
| H | -0.69808700 | 0.71375800 | 1.89521300 |
| H | 1.25192100 | 0.23468200 | 3.25274600 |
| H | -0.26943000 | -0.99585600 | 1.90428900 |
| H | 3.34256000 | -0.99237700 | 2.00061900 |
| H | 3.42982100 | 1.22281400 | 0.45823300 |
| H | 1.53284800 | 0.91195500 | 0.23128800 |

Reaction (44)

| | | | |
|---|---|---|---|
| C | 0.00000000 | 0.00000000 | 0.00000000 |
| C | 0.00000000 | 0.00000000 | 1.36900200 |
| C | 1.08478700 | 0.00000000 | 2.24951200 |
| C | 2.49344700 | 0.08335000 | 1.85646800 |
| C | 2.98459800 | 0.00626600 | 0.51314900 |
| H | -0.91406400 | 0.25693400 | -0.52230000 |
| H | 0.83236800 | -0.29908100 | -0.60941400 |
| H | -0.96778300 | 0.11550300 | 1.85361400 |
| H | 0.89817000 | -0.07336900 | 3.31414400 |
| H | 2.65910500 | -0.98768200 | 2.20254500 |

```
H            3.08689000   0.69947300   2.53445200
H            3.94695600   0.43972500   0.27563300
H            2.58290200  -0.70151700  -0.19174300
```